\documentclass[conference]{IEEEtran}
\usepackage{framed}
\usepackage[utf8]{inputenc}
\usepackage[T1]{fontenc}
\usepackage{graphicx}
\usepackage{subfig}
\usepackage{hyperref}
\usepackage[dvipsnames]{xcolor}
\usepackage{xspace}
\usepackage{enumerate}
\usepackage{enumitem}

\ifCLASSINFOpdf
\else
\fi
\begin{document}
\title{Towards Risk Modeling for Collaborative AI}

\author{\IEEEauthorblockN{Matteo Camilli \IEEEauthorrefmark{1},
Michael Felderer \IEEEauthorrefmark{2},
Andrea Giusti \IEEEauthorrefmark{3},
Dominik T. Matt \IEEEauthorrefmark{1}\IEEEauthorrefmark{3}, \\
Anna Perini \IEEEauthorrefmark{4},
Barbara Russo \IEEEauthorrefmark{1},
Angelo Susi \IEEEauthorrefmark{4}
}

\IEEEauthorblockA{ \footnotesize
\IEEEauthorrefmark{1} Faculty of Computer Science, Free University of Bozen-Bolzano, Italy \\ 
Email: \texttt{\{mcamilli,dmatt,brusso\}@unibz.it}}

\IEEEauthorblockA{ \footnotesize
\IEEEauthorrefmark{2}Dept. of Computer Science, University of Innsbruck, Austria \\
Email: \texttt{michael.felderer@uibk.ac.at}}

\IEEEauthorblockA{ \footnotesize
\IEEEauthorrefmark{3} Fraunhofer Italia Research, Bolzano, Italy \\
Email: \texttt{\{andrea.giusti,dominik.matt\}@fraunhofer.it}}

\IEEEauthorblockA{ \footnotesize
\IEEEauthorrefmark{4} Fondazione Bruno Kessler (FBK), Trento, Italy \\
Email: \texttt{\{perini,susi\}@fbk.eu}}
}

\newcommand\CAIS{CAIS\xspace}

\IEEEspecialpapernotice{\footnotesize Accepted for presentation at the 1st Workshop on AI Engineering - Software Engineering for AI (WAIN'21) \\co-located with the 43rd International Conference on Software Engineering (ICSE'21) }

\maketitle

\begin{abstract}
Collaborative AI systems aim at working together with humans in a shared space to achieve a common goal. 
This setting imposes potentially hazardous circumstances due to contacts that could harm human beings.
Thus, building such systems with strong assurances of 
%(dynamical/flexible systems) 
%while assuring
compliance with requirements  %compliance with 
domain specific standards and regulations is of greatest importance.
Challenges associated with the achievement of this goal become even more severe when such systems
rely on machine learning components rather than such as top-down 
rule-based AI. %(e.g., expert systems). 
%CAI  are gaining momentum in all domains including manufacturing industry.  
%To collaborate effectively such systems need to be able to perceive and understand their partners and act ensuring  their safety. More generally CAI, while meeting functional requirements,  need to be compliant with quality aspects and domain specific standards. This represents a key challenge in developing such systems %especially when they 
%when they rest on Machine learning (ML)  components due to their  inherent uncertainty that increases  risks of noncompliance.
%In our vision, requirements engineering, together with software and systems engineering, should contribute towards %the objective of 
%the building flexible and compliant collaborative AI
%Research here is still immature, thus further investigation is required to develop 
%by means of appropriate methods and techniques.
%including (requirements engineering, systems verification) %quality 
%has not yet developed the theoretical foundation and the associated methods to develop CAI systems with strong, ideally provable, compliance assurances minimizing the residual risks that might lead to safety, robustness, or trustworthiness issues.
In this paper, we introduce a risk modeling approach tailored to  Collaborative AI systems.
The risk model includes goals, risk events and domain specific indicators that potentially expose humans to hazards.
The risk model is then leveraged to drive assurance methods that feed in turn the risk model through insights extracted from run-time evidence. 
%The main steps of our approach include the creation of a risk model that aims at specifying goals, risk events and domain specific indicators that potentially expose humans to hazards.
%The risk model feeds verification activities that aim at increasing the level of assurance by mitigating existing risks.
%The outcome of verification is used as feedback to update the prior knowledge expressed by the risk model. 
%identify three main research directions: automated specification and management of compliance requirements, and their alignment with assurance cases; risk management; and risk-driven assurance methods. 
%Each one %of these areas yield 
%tackles challenges that currently hinder engineering processes in this context.
%the (i) different type of uncertainty that affect 
%the domain of 
%CAI engineering, (ii) the role of risk management and iterative development, as strategies to cope with such uncertainty, and (iii) the need of revising theoretical foundations of compliance requirements and compliance assurance to define appropriate engineering methods. 
%automating specification of compliance requirements, and their alignment with compliance cases, risk analysis for ML-based systems, and risk-driven assurance methods. Each research area addresses challenges that hinder engineering effective CAI systems.
Our envisioned approach is described by means of a running example in the domain of Industry 4.0, where a robotic arm endowed with a visual perception component, implemented with machine learning, collaborates with a human operator for a production-relevant task.
%A research roadmap and challenges ahead are also discussed.
%, aims at fostering further discussion on the challenges %posed by building ML-based CAIS, 
%and research directions towards
%The ultimate goal of our ongoing research is to tackle the existing challenges and provide 
%appropriate methods and tools to engineer collaborative AI systems in compliance with existing standards, norms, and regulations.

\end{abstract}


% \fi
\IEEEpeerreviewmaketitle

\section{Introduction} \label{sec:intro}

Collaborative AI systems (CAIS) work together with humans in a shared physical space to achieve common goals. 
Such systems  are required to build trust between the human and the machine by ensuring a safe interaction and not prevailing over human needs.
%
%\Anna{Collaborative AI systems (denoted as \CAIS from now on) operate in a physical space that is shared with humans who should trust  them and collaborate with them in full safe.}
%aim at working together with humans in a shared physical space to carry out tasks and achieve a common goal. 
%\CAI are gaining momentum in all domains including manufacturing industry. 
%
%Effective collaboration can be achieved by perceiving and understanding the surroundings that include human beings in a dynamic and evolving setting.
%In this context, %the \CAIS 
%%are required to build trust between the human and the machine, assuring that plans do not prevail over human needs as well as providing safety guarantees.
%In other words, \CAI need to be flexible to accommodate changing requirements while  ensuring the  satisfaction of key quality criteria for the specific application domain. 
%
%that jeopardize the compliance with %with quality aspects and 
\CAIS should be flexible to accommodate changing requirements and, at the same time, ensure the  satisfaction of key quality criteria for the specific application domain, including appropriate behavior with respect to 
social rules as well as domain specific standards %principles
and laws that are stated by certification and government bodies.
Nevertheless, such systems often run in dynamic and uncertain environments that make it difficult to provide strong assurances of acceptable behavior.
%guarantees of correctness.
This condition yields \emph{risks} of dangerous circumstances because of unfulfilled \emph{compliance requirements}. 
Systematic engineering processes to build  and assure that \CAIS are in compliance with requirements (e.g., those extracted from human-robot collaboration standards\footnote{ISO/TS 15066 (\url{https://www.iso.org/standard/62996.html}); supplementing ISO 10218‑1 (\url{https://www.iso.org/standard/51330.html}); and ISO 10218‑2 on safety (\url{https://www.iso.org/standard/41571.html}).}) %developed by international standards committees\footnote{\url{https://www.iso.org/committee/6794475.html}}) 
are still immature and require further investigation.
%Evidence supporting this claim is provided by recent 
Recent surveys report the urgent need for effective engineering processes~\cite{vogelsang2019requirements,ishikawa2019engineers} for ``intelligent'' components, such as Machine Learning (ML) models, as well as AI systems.
%In particular, unique 
%An analysis of the RE characteristics 
%for systems that include machine learning (ML) components are reported in~\cite{vogelsang2019requirements}
%A meta-model to guide the definition of structured assurance cases from requirements has been introduced in~\cite{wei2019model}, 
In particular, the work presented in~\cite{ashmore2019assuring} discusses challenges and desiderata for assurance methods of such systems.
This latter work emphasizes that existing approaches for AI systems 
%in general and \CAIS in particular 
are not linked to compliance requirements and possible risks.

% our vision
    %Moved up\Matteo{Our vision comprises \CAIS endowed with the needed flexibility in order to learn and reason on uncertain, evolving and dynamic operational ecosystems.
    %Furthermore, as engineers we should provide strong, ideally provable, assurances that learning does not lead to compliance issues.}
%

%maybe reuse later \Matteo{Here, RE processes should be guided by risks and leverage continuous feedback from empirical evidence collected at run-time, in a closed-loop with the world of interest in order to verify semantically meaningful properties through appropriate assurance methods.}

% contribution
This context brought us to reflect on how research in software and systems engineering 
%should 
could contribute to the objective of defining effective methods to build \CAIS equipped with ML components (e.g., based on neural networks). %(from now on called again \CAIS).
Our main focus is on those collaborative systems that heavily rely on ML to implement human perception skills like visual perception, speech recognition, or conversing in natural language
%to recognize and classify objects and humans in the surroundings 
under uncertain and evolving circumstances~\cite{Camilli2020SEFM}.
By collaborating with domain experts, we started to design a risk-driven approach with the objective of supporting the development of ML-equipped \CAIS, henceforth referred again to as \CAIS for the sake of simplicity.

%tailored to risk management for \CAIS.
%analyze the state of the art and the practice in light of concrete examples and then discuss open challenges and identify candidate research directions.

As part of our ongoing research~\cite{CamilliREFSQ2021}, in this paper we introduce our modeling approach by means of a running example in the context of Industry 4.0.
Essentially, we aim at dealing with risk as first class concern. %while developing \CAIS.
To do so, we propose to extend the RiskML metamodel~\cite{DBLP:conf/er/SienaMS14} to include relevant (and uncertain) characteristics of the environment that are perceived by the ML component and possibly expose execution scenarios to risk events.
Then, we envision risk-driven assurance methods that provide feedback from empirical evidence collected at run-time, in a closed-loop setting with the world of interest.
Finally, assurance methods execute high-risk scenarios and verify semantically meaningful properties to understand whether the run-time behavior of the \CAIS is acceptable.

The rest of the paper is organized as follows.
In Sect.~\ref{sec:example} we introduce a running example to put into place major concepts. Then, in Sect.~\ref{sec:approach} we elaborate on our envisioned approach.
In Sect.~\ref{sec:related-work} we discuss related work. Finally, in Sect.~\ref{sec:conclusion} we conclude the paper.

\begin{figure}[tb]
\centering
\includegraphics[width=.6\linewidth]{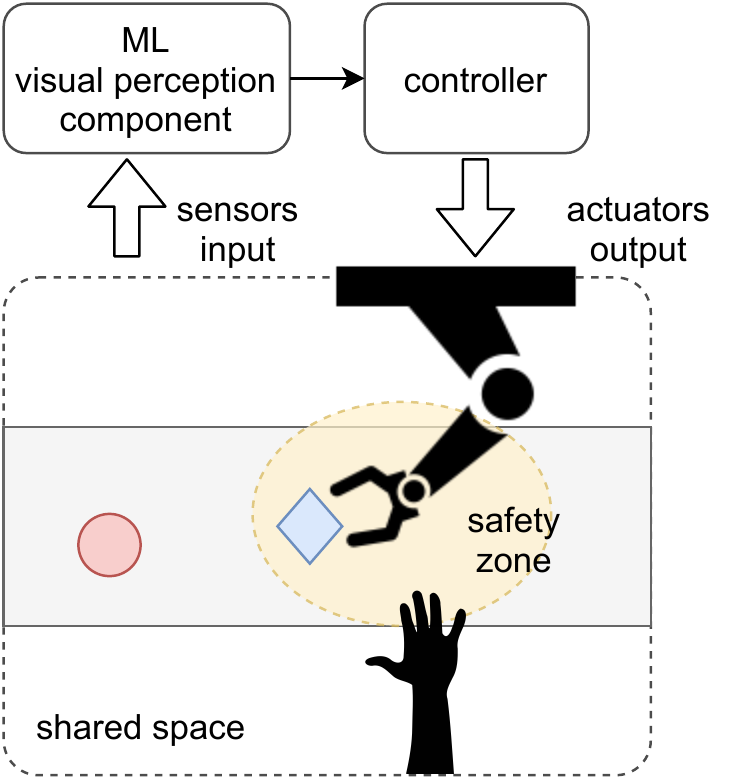}
\caption{Collaborative AI running example.} \label{fig:example}
\end{figure}

\section{A Running Example from Industry 4.0} \label{sec:example}

We illustrate our vision of the problem and the main steps of our approach by means of a running example from the Industry 4.0 domain, borrowed from~\cite{GiustiEtAl2019}, 
%~\cite{Gualtieri2020b}, 
where a robotic arm collaborates with a human operator for a production-relevant task.
%In this context, when humans and robots work together in a shared work-space, there exist high risks for human safety \cite{ISOTS15066} as well as trustworthiness\footnote{\url{https://ec.europa.eu/futurium/en/ai-alliance-consultation/guidelines}} and reliability issues.
%As stated in~\cite{Giusti2018}, when considering the standards for this domain, is particularly challenging to realize flexible automation. 
%Difficulties become particularly severe when dealing with continuous learning and adaptation to different task demands.
%
Figure~\ref{fig:example} illustrates the running example, where an automated controller of a robotic arm attempts to detect and classify objects (e.g., by color and shape) on a conveyor belt and actuates the proper movements to pick and move the object into the right bucket. The sorting skill is acquired by the robot automatically, based on human demonstrations. 
The system includes a controller, an actuated mechanical system (i.e., robotic arm), and camera sensors along with a visual perception ML component for classification. 
This ML component learns on structured heterogeneous data sources associated with \emph{features} (e.g., shape and color of an object) and yields category labels as output, where labels identify the buckets where the objects must be placed.
%These are associated to the action that the operator performs when moving the corresponding objects.
%The training and the validation is performed iteratively online by a human operator.
%
The operator collaborates with the robot in order to supervise the correct transfer of the desired sorting skill to the robot and can intervene through gestures when corrections are required.

The safety control approach of this example is implemented as in~\cite{Scalera2020}, by considering the \emph{speed and separation monitoring} approach of the standard ISO/TS 15066 for collaborative robotics.
Here, a protective separation distance between the human and the robot is checked online using \emph{safety zone}s.
The dimension of such zones is dynamically adapted based on the robot motion. %showing that promising improvement to the effectiveness of the collaboration are possible with respect to the use of static safety-zones (zones which consider worst case scenario for the robot motion). 
Fast motions of the robot can generate safety zones which may be large and therefore negatively affect the realization of collaborative operations.
%
%When handcrafting the robot motion during design of a collaborative application this aspect can be directly considered.
Assuring a successful collaboration 
%This is however not possible 
in cases in which the robot motions are learnt from humans, as in highly flexible automation, is particularly challenging~\cite{Giusti2018}.
In such scenario, we envision a novel risk mitigation process, in which  %APrelevant or better "suitable"?
suitable assurance methods are driven by an explicit risk model as described in the following.
%that elicits and quantify existing risks.
%can enable the system itself to infer risk events and adequately take decisions by e.g., learning or replicating only those skills that, most likely, will not affect the overall effectiveness of a collaboration.

%
%The \CAIS combined with the surrounding assembles a closed-loop setting where the system constantly senses and then in turn
%affects the environment.

\section{Overview of the Approach} \label{sec:approach}
In this section, we introduce our envisioned risk mitigation process for \CAIS. %Fig.~\ref{fig:process} shows an overview of the process. 
It contains risk modeling and analysis as well as risk-driven assurance, which provide feedback to each other in order to mitigate known risks below a defined threshold.
%
%\begin{figure}[htb]
%\centering
%\includegraphics[width=.6\linewidth]{process.pdf}
%\caption{Iterative risk mitigation process.} %\label{fig:process}
%\caption{Cycling process of risk Modeling \& analysis and risk-driven assurance}
%\end{figure}
%
\begin{figure*}[tb]
\centering
\includegraphics[width=.75\linewidth]{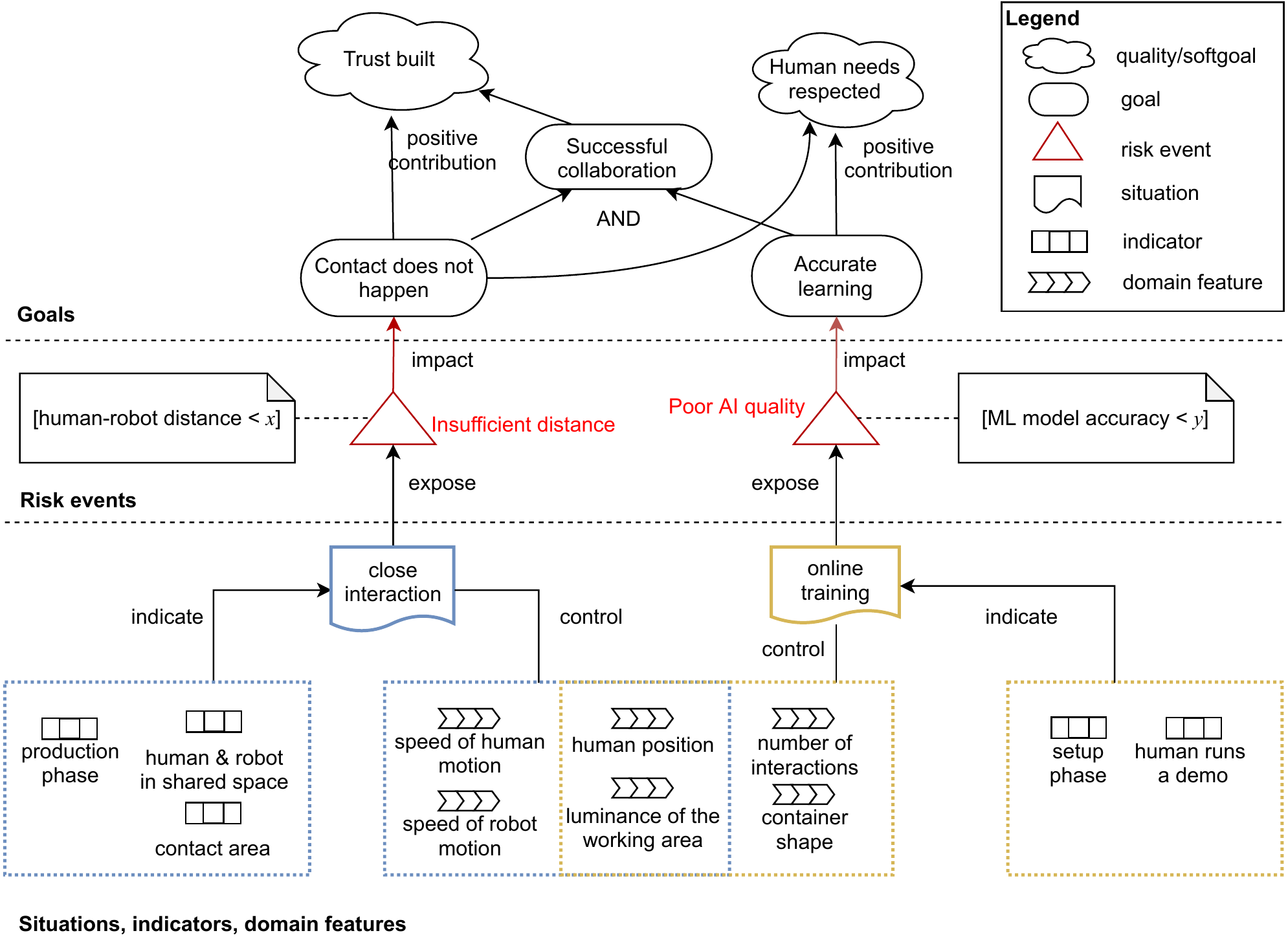}
\caption{RiskML model extract of our running example.}
\label{fig:model}
\end{figure*}
\subsection{Risk Modeling \& Analysis}
%
%\textbf{short overview of RiskML (Angelo)}, \textbf{RiskML for AI example (Angelo)}\\
The risk modeling \& analysis phase is based on the RiskML language~\cite{DBLP:conf/er/SienaMS14} that we extend with the notion of \textit{domain feature} to model aspects of the environment that can be perceived directly by AI components of the system under consideration. 
Figure~\ref{fig:model} shows a model of goals and risks for the collaborative learning example introduced in Sect.~\ref{sec:example} described using the extended RiskML notation. In the model, the concept of \textit{Situation} is used to model the circumstances
%state of affairs 
under which a certain risk is possible, such as, for example, the situation \textsf{close interaction}.
An \textit{Event} models a change in the circumstances, with a potential negative impact on the goals of the system (e.g., the event \textsf{insufficient distance}). Events have a certain (negative) impact on the affected assets, and they occur with a certain likelihood. The notion of \textit{Goal} models the desire of a stakeholder in obtaining or maintaining some business value, such as, for example, the goal \textsf{contact does not happen}.
The concept of an \textit{Indicator} embodies a characteristic of a situation, such as, for example, \textsf{human and robot are in a shared space}, as indicator of the situation \textsf{close interaction}.

As mentioned, we finally enrich the RiskML meta-model with the concept of a \textit{Domain Feature}, which we introduce to capture relevant and quantifiable characteristics of the environment for a given situation.
For instance, the ML visual perception component in our example tries to understand what happens in the shared space where a number of aspects are uncertain and vary over time.
Such uncertain aspects, influencing the perception capability, shall be explicitly modeled to define the space of possible environment configurations under which the \CAIS works. 
For example, the \textsf{close interaction} situation can lead to risks depending on the \textsf{luminance of the working area} and the \textsf{speed of the human motion}.
Thus, domain features yield semantically meaningful inputs for an AI component whose
actual values ranges in given domains determined by the context. %specified by the modeler.
For instance, the luminance can assume either discrete or continuous values in a given interval that depends on the assumptions about the target production environment.
%
%Both Indicators and Domain Features can contribute to characterise situations in the model and take values from the environment. For instance the \texttt{luminance of the working area} can take values in a range of possible luminance values in the given domain described via a discrete or continuous set of values or via an equation. 
%Focusing the attention to the left part of the model in Fig.~\ref{fig:model}, and starting from the situation \texttt{Close Interaction}, we have that it is characterized by a set of indicators, \texttt{production phase}, \texttt{human and robot move together in shared space}, and \texttt{contact area} and a set of domain features as input to the AI component, \texttt{speed of human motion}, \texttt{speed of robot motion}, \texttt{human position}, and \texttt{luminance of the working area}.

The model extract in Fig.~\ref{fig:model} shows two possible situations: \textsf{close interaction} on the left and \textsf{online training} on the right.
The situation \textsf{close interaction} \textit{expose}s the system to \textsf{insufficient distance}, i.e., a risk event that has an \textit{impact} on one of the goals of the system, i.e., \textsf{contact does not happen} in this case. The risk can be quantified by verifying the human-robot distance against a given threshold $x$. This impact can compromise the satisfaction of the higher level goals \textsf{Successful collaboration}, compromising in turn the general goal of \textsf{Trust built}.
The \textsf{online training} situation
\textit{expose}s the system to \textsf{poor AI quality}, i.e., the risk event that has an \textit{impact} on the goal \textsf{accurate learning}, which in turn impacts the general goal \textsf{Human needs respected}.

%
%The overall risk expresses a lack of knowledge about some happening and what could be the consequences and is a composition of situations, in which risk events potentially occur, an event whose occurrence impacts on goals, and the goal, which suffers an impact if the event occurs. 
%
The quantification of risks for a specific situation (e.g., \textsf{close interaction}) is performed on the basis of indicators and domain features during risk analysis. The thresholds and probability distributions of values defined for some (or a combination) indicators and features determine the likelihood of the occurrence of (one or more) risk event(s). As an example the situation \textsf{close interaction} aggregates the values of the different indicators (e.g., \textsf{human and robot in the shared space} = $\mathtt{True}$) and from features (e.g., \textsf{speed of the human motion} in and \textsf{speed of the robot motion} in the continuous ranges $[0.0,2.0]$ m/s and $[0.5, 2.0]$ m/s, respectively), so recognizing a situation of an active collaboration %between the human and the robot 
that may become risky (occurrence of a risk event) when the speed and directions of the human and robot movements lead to violate the safety zone (i.e., the distance became less then $x$). In addition, the impact may explicitly be determined or it is considered high for all explicitly modeled events and situations.

The risk analysis drives the assurance process.
It is important to point out that thresholds and value distributions are adapted during the assurance, which is described in the next subsection.

\subsection{Risk-driven Assurance}

Our approach maps \emph{Situation}s in the model to compelling \emph{assurance case}s~\cite{ashmore2019assuring} that expose the system to \emph{risk event}s.
Assurance cases are built according to indicators and then executed in order to provide evidence that the run-time behavior is acceptable under such circumstances.
In particular, given the inherent infeasibility of ``correct by design'' AI systems~\cite{DreossiCAV2019}, we aim at providing run-time assurances by running the \CAIS embedded in a closed-loop setting with a simulated environment (e.g., using \textsc{Gazebo}\footnote{\url{http://gazebosim.org/}} or \textsc{Webots}\footnote{\url{http://www.cyberbotics.com/}}).
Thus, an assurance case is simulated multiple times under a number of alternative environment configurations.
Such configurations are sampled from the \emph{search space} defined by the set of domain features attached to the corresponding situation.
Each individual simulation, can lead to a risk event that is detected by constant monitoring of measurable phenomena, depending on the quantitative \emph{constraint}s attached to risk events.
For instance, the event \textsf{insufficient distance} occurs when the distance between the human and the robot is less than $x$, with $x$ determined by the dynamic safety zone, as shown in Fig.~\ref{fig:example}.

Once the domain and range of the relevant domain features are defined by the modeler, we can automatically search for executions that violate model constraints.
Since exhaustive enumeration of all possible environment conditions is in general too expensive, we leverage on meta-heuristic optimizing search techniques~\cite{10.5555/3001602} to guide the sampling towards violations (i.e., minimization of the human-robot distance in our example).
Violations of model constraints yield execution scenarios in which the \CAIS does not meet one or more goals according to the risk model.
In this case, the actual values assigned to domain features can be used to explain the undesired behavior of the system and drive the retraining of the ML component.
For instance, low luminance values might lead to decrease the ability to correctly classify the human operator in specific locations of the working area. 
%and thus, increase the likelihood of the event $\mathtt{insufficient~distance}$.

As suggested by authors in~\cite{DreossiCAV2019}, in addition to meta-heuristic optimization, \emph{importance sampling} techniques (e.g., Bayesian optimization sampling) can be used to update a prior sampling distribution through run-time evidence provided by the execution of the assurance cases.
The outcome of this process is a distribution called \emph{infill sampling criterion} which encourages the important values for the domain features.
The notion of importance here depends on the target situation exercised by an assurance case.
For instance, considering the \textsf{insufficient distance}, important values are likely to minimize the human-robot distance.
Such a distribution provides a meaningful feedback to the risk analysis as well as the risk model.
For example, sensitivity analysis can be applied to classify the domain features in order to detect the most important factors associated with risks.
Furthermore, the highest density region of the infill sampling criterion provides the modeler with the \emph{credible interval}s of the important values.
Loosely, the larger the intervals the higher the likelihood of risks in a given situation.
Hence, the magnitude of the credible intervals provides the basis for estimating the probability associated with risk events and thus, close the feedback loop.

\section{Related Work} \label{sec:related-work}
%
%approaches
%, may exceed what is considered in current standards. Difficulties 
%arise from frequently changing environments and potentially unknown \textit{a-priori} robot motions. These issues 
Challenges in flexible automation become particularly severe when robotic systems swiftly adapt to different tasks by learning from humans using ML components~\cite{Billard2016learning,Giusti2018}.
%In fact, existing standards do not specifically refer to \CAIS able to mimic aspects of human intelligence through ML.
%Social rules and guidelines naturally emerge from the field that is constantly evolving because of new enabling technologies.
%To address this issue, several initiatives from international organizations, including the aforementioned international standards committee for human-robot collaboration, started in the last years.
Understanding how to properly assure compliance requirements resulting from standards is still an open issue~\cite{vogelsang2019requirements,ishikawa2019engineers,ashmore2019assuring}.

Risk management have been playing an important role in software and systems engineering processes.
The results of risk analysis are typically applied to drive requirements engineering~\cite{cailliau2012probabilistic} and testing~\cite{felderer2014risk} activities.
However, risk management processes in the context of \CAIS is immature and requires further investigation. 
The work presented in~\cite{debrusk2018risk} discusses the risk of susceptible to unintended biases of ML, whereas in~\cite{foidl2019risk} authors consider risks for data validation in ML systems. 
%An approach supporting risk management and prioritisation activity in a semi-automated way has been introduced in~\cite{AvesaniPSS15}.
%It merges the capability to exploit existing risk-related information in a given organisation with an automated ranking of the requirements with respect to the level of risk the decision-maker estimates for them. %The approach combines knowledge about risks assessment techniques and Machine Learning to enable an active intervention of human evaluators.
A common risk assessment approach for collaborative robotics applications is based on ISO 12100~\cite{iso201012100}.
%as mentioned in ~\cite{ISOTS15066}.
An example of risk assessment process based on this standard is reported in~\cite{Bjoern2016example}. 
%in the context of collaborative assembly in case of potential contacts between the robot and the operator.
This latter approach does not take into account AI components embedded into robotic systems.
%takes into account relevant uses cases and possible hazardous contact events.
%The associated risk based on severity and probability are then evaluated in order to decide whether the risk is acceptable.

As stated in~\cite{ashmore2019assuring}, assurance methods in the lifecycle of AI systems need a step change.
%are concerned with the provision of provable evidence that either the ``intelligent'' component or the whole system will continue to satisfy requirements when exposed to the production environment for its intended application.
Offline approaches are usually considered more optimistic compared to online approaches~\cite{UlHaq2020Comparing}.
In fact, the online setting aims at generating execution scenarios that might cause violations of requirements by accounting for the closed-loop behavior of AI systems~\cite{Gambi2019ISSTA}.
As advocated in~\cite{DreossiCAV2019}, more accurate specifications can be achieved only by considering the execution context (i.e., capturing the model of the environment).
Thus, effective assurance methods for \CAIS
%as well as AI systems in general, 
shall should analyze the ML components online, along with their context so that semantically meaningful properties can be verified~\cite{DreossiCAV2019}.

\section{Conclusion and Future Directions} \label{sec:conclusion}

The limited availability of systematic and effective approaches to compliance assurance for \CAIS, currently prevents
%limits 
their widespread adoption in different domains such as industrial manufacturing.
Domain experts advocate the need of effective risk-driven approaches to build \CAIS. 
In this paper we introduced a high level overview on our modeling approach that considers risk as primary concern with 
the ultimate objective of building \CAIS with strong, ideally provable compliance assurances. %by mitigating existing risks.
%Major challenges that hinder our ultimate goal include: 
%uncertain environment, need of standards, partial and evolving specifications, and top-down/bottom-up duality.
As part of our ongoing research, we plan to apply our modeling approach as well as the risk-driven assurance process to build \CAIS in real world settings in order to assess applicability and cost-effectiveness by means of empirical evidence.

\bibliographystyle{IEEEtran}
\bibliography{references}

\end{document}